\documentstyle [12pt] {article}
\begin{document}

\centerline{\bf Density perturbations in Kaluza--Klein theories during a de
Sitter phase}

\vskip24pt
\baselineskip=10pt
\centerline{J\'ulio C\'esar Fabris$^{\diamond}$ and Mairi
Sakellariadou$^{\star}$}
\vskip10pt
\centerline{\it $^{\diamond}$Departamento de F\'{\i}sica}
\centerline{\it Universidade Federal do Esp\'{\i}rito Santo - UFES,
Av. Fernando Ferrari}
\centerline{\it s/n, CEP 29060-900, Vit\'oria-ES, Brazil}
\vskip 11pt
\centerline{\it $^{\star}$D\'epartement de Physique Th\'eorique}
\centerline{\it Universit\'e de Gen\`eve}
\centerline{\it 24 quai Ernest Ansermet, CH-1211 Gen\`eve 4, Switzerland}
\vskip 44pt
\begin{abstract}
\vskip 10pt
In the context of Kaluza-Klein theories, we consider a model in which the
universe is filled with a perfect fluid described by a barotropic 
equation of
state.  An analysis of density perturbations employing the synchronous gauge
shows that there are cases where these perturbations have an exponential
growth during a de Sitter phase evolution in the external space.
\vspace{1cm}
\end{abstract}
\vspace{1cm}

\section{Introduction}
One of the main still open questions within the realm of modern cosmology is
the origin of the primordial density fluctuations which played the role of
seeds for structure formation. Within the framework of gravitational 
instability, the two alternative scenarios which could lead
to the generation of perturbations are either the amplification of quantum
fluctuations during an inflationary era \cite{stein}, or fluctuations seeded
by topological defects \cite{kibble}.  Both inflation and topological defect 
models lead to approximately scale invariant Harrison-Zel'dovich spectra on 
large angular scales, thus both are consistent with the COBE-DMR 
results \cite{cobe}.  The characteristic features of the Doppler peaks in the 
angular power spectrum of the cosmic microwave background, might be an 
important
discriminating tool between inflationary fluctuations and those produced by
topological defects \cite{doppler}.
The dynamics of the initial density
fluctuations has been studied by a large number of authors, either in the
context of general relativity, or in multi-dimensional theories.

In a generic inflationary model, the inflaton field $\phi$ experiences 
quantum fluctuations on all scales smaller than the effective particle 
horizon at the onset of inflation.  During the era of exponential expansion, 
the scales inflate outside the horizon  and therefore quantum fluctuations 
freeze in as classical fluctuations of $\phi$, leading to generation of 
energy density perturbations.  The amplitude of the density perturbations 
of a given scale $\lambda$ when they cross back the horizon at the end of
inflation is \cite{turner}
\begin{equation}
\left({\delta \rho \over \rho}\right)_{hor} \sim 
\left({H^2\over \dot\phi}\right)_{\lambda} \ .
\end{equation}
Thus, through a linear mechanism one obtains a spectrum of density 
perturbations of constant amplitude, which under certain choices lead to the 
required amplitude of initial fluctuations, sufficient to generate structure 
formation without distorting the isotropy of the cosmic microwave 
background.  

For a number of authors, inflation is plugged with difficulties (e.g., 
potential fine tunning of coupling constants in order to achieve the required 
amplitude of density perturbations) and therefore we believe that it is
important to investigate alternative scenarios.  As such, one can study
topological defect models (which will not be addressed here) or the 
growth of density perturbations in multi-dimensional theories, in 
particular, during a de Sitter phase.  
One may wonder whether in order to get non-zero density perturbations during 
a de Sitter phase while remaining within the framework of general relativity, 
one should either consider theories where the energy momentum is not
conserved, or to remove the condition of a perfect fluid.
We will sketch why none of these attemps work and therefore motivate 
our choice to work with multi-dimensional theories. 

A scalar-tensor theory of gravity which involved a non-conserved
energy momentum tensor was proposed by Jordan \cite{jor}, but it was
criticized \cite{bondi}, basically because it fails to successfully
incorporate non-relativistic matter.  A study of the evolution of scalar
perturbations in a universe filled with a viscuous fluid, in the framework
of general relativity, shows that the density perturbations decay (see,
appendix A).  So, to
investigate whether scalar perturbations can grow during a de Sitter phase,
we will consider a universe filled with a perfect fluid described by a
barotropic equation of state, in the context of Kaluza-Klein \cite{kk}
theories, where energy momentum conservation allows density perturbations
provided the de Sitter phase characterizes only the four-dimensional subspace.
  
In this work we are only dealing with the evolution of scalar perturbations,
without examining their origin. In the context of inflation, for example,
initial perturbations are generated by quantum fluctuations of the inflaton 
field. 
We will perform our analysis employing the synchronous gauge, while
keeping in mind the residual coordinate freedom in order to finally keep
only the physical modes.	

Our presentation is organized as follows:  In section 2 we first set 
up the problem, specify the notation, write down the background equations and 
the energy momentum tensor. In section 3 we introduce linear perturbations and
derive the set of equations that describe the evolution of scalar
perturbations in the synchronous gauge.  In section 4 we  solve the set of
equations for two  particular cases, where the external space is in a de
Sitter phase.  While our choices for these cases
may be rather special, our aim here is only to show that it is possible to
obtain growth of perturbations; the issue of how generic this situation is,
is left for future work.  We will show that, within the framework of
multi--dimensional theories, we can indeed find at least one
example where there is growth of scalar perturbations, in particular with an
exponential behaviour, during a de Sitter phase.  We summarize our results in 
section 5.  We close with two appendices; in the first one we discuss density 
perturbations in a viscous
universe, while in the second one we analyze the problem of the residual
coordinate freedom.
\section{The model}
The Lagrangian density in a $n$-dimensional spacetime, coupling
gravity to ordinary matter, is
\begin{equation}
{\it L} = \frac{1}{16\pi G}\sqrt{- g} R - {\it L_m} \ ,
\end{equation}
where all quantities are defined in the multi--dimensional spacetime. 
{}From this expression, we can deduce the field equations in higher
dimensions
\begin{eqnarray}
\label{fea}
R_{AB} - \frac{1}{2}g_{AB}R &=& 8\pi GT_{AB} \ , \\
\label{feb}
{T^{AB}}_{;B} &=& 0 \ .
\end{eqnarray}
We consider that each spatial section is divided into two spaces, the
external one of dimension $d_1$ and the internal one of dimension $d_2$, such
that $n = 1 + d_1 + d_2$.  We assume the metric to be of the form
\begin{equation}
ds^2 = - dt^2 + a(t)^2\gamma_{ij}dx^idx^j + b(t)^2\gamma_{ab}dx^adx^b\ ,
\end{equation}
where $\gamma _{ij}$ is the external metric and $\gamma _{ab}$ the internal
one, and both spatial sections have a constant curvature, not necessarily
zero. Normalizing the curvature, we write $k_1 = -1$, $0$ or $1$ and
$k_2 = -1$, $0$ or $1$ for the curvatures of the spaces with dimensions
$d_1$ and $d_2$ respectively.\\
We consider an anisotropic energy momentum tensor, with different
pressures in each subspace. Thus, it takes the form
\begin{eqnarray}
T^{AB} &=& (\rho + \bar p)u^Au^B - p_1u^iu^{(A}\delta^{B)}_i -
p_2u^au^{(A}\delta^{B)}_a - \nonumber \\
& &  \bar pg^{AB} +
\frac{p_1}{2}g^{i(A}\delta^{B)}_i + \frac{p_2}{2}g^{a(A}\delta^{B)}_a \ ,
\end{eqnarray}
where
\begin{equation}
\bar p = \frac{1}{2}(p_1 + p_2)\ \ \ ; \ \ \ 
u^{(A}u^{B)} = \frac{1}{2}(u^Au^B + u^Bu^A)\ , \nonumber \\
\end{equation}
with  $i = 1, \ldots, d_1$; $a = 1, \ldots, d_2$. \\
We consider that both pressures have a barotropic equation of state: 
\begin{equation}
p_1 = \alpha_1\rho \ \ \ ;  \ \ \ p_2 = \alpha_2\rho\ . \nonumber \\
\end{equation}
The differential equations we obtain from Eqs.~(\ref{fea}) and
(\ref{feb}), relating scale factors $a(t)$, $b(t)$ and $\rho$ are
\begin{eqnarray}
d_1\frac{\ddot a}{a} + d_2\frac{\ddot b}{b} &=& \nonumber \\
 - \frac{8\pi G}{d_1+d_2-1}
\biggr[d_1(1 + \alpha_1) + d_2(1 + \alpha_2) - 2\biggl]\rho &&\ ;\label{eq1}
\\
\frac{\ddot a}{a} + (d_1 - 1)(\frac{\dot a}{a})^2 +
d_2\frac{\dot a}{a}\frac{\dot b}{b} + (d_1 - 1)\frac{k_1}{a^2} &=& \nonumber
\\
\frac{8\pi G}{d_1+d_2-1}\biggr[1 - \alpha_1 + d_2(\alpha_1 - \alpha_2)
\biggl]\rho &&\ ;\label{eq2}\\
\frac{\ddot b}{b} + (d_2 - 1)(\frac{\dot b}{b})^2 + d_1\frac{\dot a}{a}
\frac{\dot b}{b} + (d_2 - 1)\frac{k_2}{b^2} &=& \nonumber \\
\frac{8\pi G}{d_1+d_2-1}\biggr[1 - \alpha_2 + d_1(\alpha_2 - \alpha_1)
\biggl]\rho &&\ ;\label{eq3}\\
\dot\rho + d_1(1 + \alpha_1)\frac{\dot a}{a}\rho +
d_2(1 + \alpha_2)\frac{\dot b}{b}\rho &=& 0 \ .\label{eq4}
\end{eqnarray}
Some special solutions for these equations were found by Sahdev \cite{sah}.
\section{Perturbation equations}
\vspace{0.5cm}
We now proceed with the perturbative level. We introduce in Eqs.~(\ref{fea})
and (\ref{feb}) the quantities 
\begin{equation}
\tilde g_{AB} ={}^0{g}_{AB} + h_{AB}\ \ \ ,
\ \ \ \tilde\rho = {}^0\rho + \delta\rho \ \ \ , 
\ \ \ \tilde p = {}^0{p} + \delta p\ , \nonumber \\
\end{equation}
where
${}^0{g}_{AB}$, ${}^0\rho$ and ${}^0{p}$ represent the background solutions
while $ h_{AB}$, $\delta\rho$ and $\delta p$ are small perturbations around
them. We will also impose that the perturbations behave spatially like plane
waves. Due to the anisotropy of the  space, the equations take a tractable
form only if the wave is defined in just one space. We consider that
all perturbed functions depend only on the coordinate of the external space,
{\it i.e.,} $\delta(x_i,t) =\delta(t)\exp(i\vec q \cdot \vec x)$, where $x_i$
denote the coordinates in the space of dimension $d_1$.

The equations are invariant under an infinitesimal coordinate transformation
$x^A \rightarrow x^A + \chi^A$. So, we have the freedom of imposing a
coordinate condition. We will impose the synchronous coordinate condition
\cite{lk},
\begin{equation}
h_{A0} = 0 \ .
\end{equation}
As it is well known, there is yet a residual coordinate freedom \cite{press},
represented by a coordinate transformation that preserves the coordinate
condition. This can lead to non-physical modes in the final solutions. We
will discuss later this problem, establishing the criterium to eliminate
them (see, appendix B).\\
We define $h = h_k^{\ k}/a^2$, $H = h_a^{\ a}/b^2$, $\Delta 
=\delta\rho/\rho$.
After a long but straightforward calculation, we obtain from Eqs.~(\ref{eq1})
-- (\ref{eq4}) the following  system of differential equations
\begin{eqnarray}
\ddot h + 2\frac{\dot a}{a}\dot h + \ddot H + 2\frac{\dot b}{b}\dot H
&=& 2\biggr(d_1\frac{\ddot a}{a} + d_2\frac{\ddot b}{b}\biggl)\Delta
\label{de1}\ ; \\
\ddot H + (d_1\frac{\dot a}{a} + 2d_2\frac{\dot b}{b})\dot H
+ \biggr[\frac{q^2}{a^2} - 2(d_2-1)\frac{k_2}{b^2}\biggl]H &=&
- d_2\frac{\dot b}{b}\dot h + \nonumber \\
2d_2\biggr[\frac{\ddot b}{b} +
(d_2-1)(\frac{\dot b}{b})^2 + d_1\frac{\dot a}{a}\frac{\dot b}{b}
+ (d_2-1)\frac{k_2}{b^2}\biggl]\Delta\label{de2} &&\ ; \\
\dot\Delta + (1 + \alpha_1){\delta u^i}_{,i} + \frac{1}{2}(1 + \alpha_1)
\dot h + \frac{1}{2}(1 + \alpha_2)\dot H &=& 0\label{de3} \ ; \\
(1 + \alpha_1)\delta\dot u^i + (1 + \alpha_1)\biggr[(2 - d_1\alpha_1)
\frac{\dot a}{a} - d_2\alpha_2\frac{\dot b}{b}\biggl]\delta u^i
&=&\nonumber \\
 - \frac{1}{2}(\alpha_1 - \alpha_2)H^{,i} - \alpha_1\Delta^{,i}
\label{de4}  &&\ .
\end{eqnarray}
These equations describing the evolution of scalar perturbations, are too
complicated to be solved in the general case.  In the next section, we 
will solve them in two particular cases, where the equation of state in 
the external space is $p_1=-\rho$.
\section{Specific cases}
Here, we will solve the system of differential equations describing the 
evolution of density perturbations in the framework of multi--dimensional
theories with $\alpha =-1$, for the following particular cases:
\vspace{0.5cm}\\
(i)    Both the external and the internal spaces
have scale  factors with a power-law behaviour, $a\propto t^r$ ($r>0$) and
$b\propto t^s$, respectively. The internal space is also flat.  To get a
non-vanishing $\delta\rho/\rho$, the set of equations describing the
evolution of density perturbations requires $\alpha_2\ne -1$.  On the other
hand, the background equations and the energy conservation equation imply
\begin{eqnarray}
s&=&{2\over d_2(1+\alpha_2)}~;\\
r&=&{d_2 s(s-1)\over d_2s+d_1-1}~.
\end{eqnarray}
One can easily check that Eqs.~(\ref{de1}) -- (\ref{de4}) imply
\begin{equation}
\Delta=-{1\over 2}(1+\alpha_2)H + const~\label{d1}
\end{equation}
and lead to the following third order equation for $H$ :
\begin{eqnarray}
H''' + H''  {1\over \eta(1-r)}\Bigl[d_1r+d_2s-r+1\Bigr] &&\nonumber \\
+H'\left[q^2+{1\over
[\eta(1-r)]^2}\{d_1(r^2+2r)+d_2s(2+3r-2s)-r^2-2\}\right] &&\nonumber \\
+H\left[{q^2\over \eta(1-r)}+{2\over [\eta
(1-r)]^3}\{d_1r^2+d_2s(2r-s)-2r+1\} \right]&=&0\label{b1},
\end{eqnarray}
where primes denote partial derivatives w.r.t.\ conformal time $\eta$, 
defined
by $ad\eta =dt$.  After some manipulations, we obtain that, as far as time
evolution is concerned, Eq.~(\ref{b1}) reduces to the second order equation
\begin{eqnarray}
{d^2g\over d\tau^2}+{1\over \tau}{dg\over d\tau} +g\biggl[ 1-{1\over \tau^2}
{1\over 4(1-r)^2}
\{1+r^2(d_1^2-4d_1+4) &&\nonumber \\ +s^2d_2(d_2+8)-r(6d_1-4)-6d_2s
+2rsd_2(d_1-6)\}\biggr] &=& 0~\label{b2},
\end{eqnarray}
where
\begin{eqnarray}
g&=&\Bigl[H \eta^{1/ (1-r)}\Bigr]' \tau^{\gamma}~; \\
\gamma&=&{ 3-d_1r-d_2s\over 2(r-1)}~; \\
\tau^2&=&q^2\eta^2~.
\end{eqnarray}
Equation (\ref{b2}) is a Bessel equation, whose solutions are Bessel 
functions
of order
\begin{equation}
\nu={d_2^4s^4+2d_2^3s^3(s+2)+d_2^2s^3(s+4)+8d_2^2s^2+4d_2s(s+2)+4\over
4[d_2s(2-s)+2]^2}~.
\end{equation}
This expression  simplifies a lot
if $d_1=3$ and $d_2s=-1$, which means that the external space grows like
$t^r$ with $r>1$, while the internal one goes like $t^s$ where $s<0$.
In that case,
\begin{equation}
\nu={r\over 2(1-r)}=-{1+d_2\over 2}~.
\end{equation}
The general solution of Eq.\ (\ref{b2}) is
\begin{equation}
g(\tau)=uJ_\nu(\tau)+vJ_{-\nu}(\tau)\ ,
\end{equation}
where $u$ and $v$ are arbitrary constants.  For the sign in the argument of
the Bessel functions we choose $\tau =+q\eta $.\\
\par
In terms of the density contrast, we find
\begin{equation}
\Delta = -\frac{1}{2}(1+\alpha_2){\eta^\frac{1}{r-1}}q^{\gamma}\int \eta^
{-\gamma}\biggr(uJ_\nu(q\eta) +
vJ_{-\nu}(q\eta)\biggl)d\eta .
\end{equation}
For the particular case we are considering here, and re-expressing the
solutions in terms
of the cosmic time $t$, we find the asymptotic behaviour for $\Delta$,
for $t \rightarrow 0$ ($\eta \rightarrow \infty$) and $t
\rightarrow \infty$ ($\eta \rightarrow 0$), namely
\begin{eqnarray}
&t& \rightarrow 0 \Longrightarrow \Delta \rightarrow
t^{-\frac{d_2 + 2}{2d_2}}(c_1\cos t^{-\frac{1}{d_2}} +
c_2\sin t^{-\frac{1}{d_2}}), \\
&t& \rightarrow \infty \Longrightarrow \Delta \rightarrow
t^{\frac{-d_2 - 5 \pm 2(d_2 + 1)}{2d_2}} .
\end{eqnarray}
So, initially the density contrast has an oscillatory behaviour
with decreasing amplitude, while asymptotically it tends only to
decreasing modes.
\vspace{0.5cm}\\
(ii) The external space is a de Sitter flat space filled with a perfect fluid
described by $p_1=-\rho$, and its scale factor goes like $a\propto e^{rt}$,
where $r$ is a constant.  The internal space has dimensions $d_2>3$, constant
non-zero curvature and constant scale factor.  The energy conservation
equation implies $\rho=const$ and the background equations lead to
$d_2<2/(1+\alpha_2)$.  To get a non-zero $\delta\rho/\rho$, the perturbation
equations require $\alpha_2 \ne -1$.  The system of equations describing
the evolution of the scalar perturbations simplifies to :
\begin{eqnarray}
\ddot h + 2 r \dot h + \ddot H &=&6r^2 \Delta\label{s1}~;\\
\ddot H + d_1r\dot H-\left[2(d_2-1){k_2\over b_2}-{q^2\over
e^{2rt}}\right]H&=&
(d_2-1){k_2\over b^2}\Delta d_2\label{s2}~;\\
\dot{\Delta } +{1\over 2}(1+\alpha_2)\dot H&=&0\label{s3}~;\\
 \Delta _{,i} +{1\over 2}(1+\alpha_2)H_{,i} &=&0~\label{s4}.
\end{eqnarray}
Clearly, Eqs. ~(\ref{s2}) -- (\ref{s4}) again imply
\begin{equation}
\Delta=-{1\over 2}(1+\alpha_2)H + const~.
\end{equation}
To get a growing solution of Eq.~(\ref{s2}) we find the same
requirement as the one imposed by the energy conservation equation, namely
\begin{equation}
d_2<{2\over 1+\alpha_2}~.\nonumber
\end{equation}
Then, the solution for the density perturbation is
\begin{equation}
\Delta = \tilde {\tau}^{\frac{d_1}{2}}\biggr(c_1J_\nu(q\tilde {\tau}) + 
c_2J_{-\nu}(q\tilde {\tau})\biggl)
\end{equation}
where $\tilde {\tau} = - \frac{e^{-rt}}{a_0r}$ and
\begin{equation}
\nu = \sqrt{\frac{{d_1}^2}{2} + d_1 + 1 + \alpha_2(d_1 - 1)}~.
\end{equation}
The asymptotic behaviour is easily obtained. For small $t$, large
$\tilde {\tau}$,
\begin{equation}
\Delta = e^{-\frac{r}{2}(d_1-1)t}cos(\beta e^{-rt} + \delta)~,
\end{equation}
where $\beta$ and $\delta$ are constants.
For large $t$, small $\tilde {\tau}$, we obtain a mode which exhibts 
exponential growth,
\begin{equation}
\Delta \sim e^{r(-\frac{d_1}{2} + \nu)t}.
\end{equation}
So, the density perturbations exhibt initially an oscillation with
decreasing amplitude and then evolve towards a behaviour characterized by 
an exponential growth.
\section{Conclusions}
One of the most interesting and important issues within modern cosmology,
is undoubtfully the origin of large scale structure.  Within the framework of
gravitational instability, there are two classes of theories attempting to
answer the question of structure formation: inflationary models and 
topological defect scenarios.  The cosmic microwave background anisotropies,
attempting to bridge between theoretical models and observational data,
might support or rule out one of the two families of structure formation. 
In the context of inflation, quantum fluctuations of the scalar field give rise
to density perturbations only once inflation ended and the perturbations 
re-entered the horizon. In this work, we are mainly interested in a model 
where density perturbations can grow during the inflationary era.
 We are working in the context of Kaluza-Klein theories, where the external
space inflates. 

We employ some special solutions of multi-dimensional theories,
and study more closely two particular cases, one with power-law
behaviour of the scale factor, and one with exponential-type behaviour.
The first one leads only to asymptotical decaying modes, while the second 
one, in spite of an oscillatory decreasing initial behaviour, gives 
asympotically an exponential growth of density perturbations.
This indicates that such multi-dimensional theories may furnish sufficiently 
large perturbations in the beginning of the radiation dominated era, thus 
they may provide an alternative picture. However, one should 
keep in mind that a further analysis, for example the determination of 
the amplitude of quantum fluctuations, which may allow a  confrontation with 
observational data, is required before a final statement is reached. 

\begin{appendix}
\section*{Appendices}
\vspace{0.5cm}
\section{Perturbations in a viscous universe}
\vspace{0.5cm}
Another way of obtaining a growth of density perturbations during an 
inflationary phase,
is to use an imperfect fluid in the energy-momentum tensor. This can be
achieved, for example, by introducing viscosity. Here we consider bulk
viscosity in the context of the four-dimensional general relativity. The
bulk viscosity effects can be deduced just by replacing $p \mapsto p -
\chi(\rho)\Theta$ in the energy-momentum tensor, Eq.~(\ref{feb}), where
$\Theta = {u^\mu}_{;\mu}$. We  consider a barotropic equation of state, 
$p =
\alpha\rho$.\\ The field equations are
\begin{eqnarray}
3(\frac{\dot a}{a})^2 &=& 8\pi G\rho \ ; \\
2\frac{\ddot a}{a} + (\frac{\dot a}{a})^2 &=& - 8\pi G\alpha\rho +
24\pi G\chi(\rho)\frac{\dot a}{a} \ ; \\
\dot\rho + 3(1 + \alpha)\rho &=& 9\chi(\frac{\dot a}{a})^2 \ .
\end{eqnarray}
If $\chi(\rho) = \chi_{_0} = const$, we have the simple solution
$a \propto e^{Ht}$, where $H = (1+\alpha)\rho/3\chi_{_0}$ and $\rho = const$.
If on the other hand, $\chi(\rho) =\chi_{_0}\rho$, we have the solutions 
found
by Murphy \cite{mur}, which reduce either to the previous one in the 
limit $t
\rightarrow - \infty$, or to the flat perfect fluid solution in the limit $t
\rightarrow \infty$.

If we perturb Einstein's equations in the presence of viscosity,
we find the following coupled differential equations
\begin{eqnarray}
\ddot h + 2\frac{\dot a}{a}\dot h - 12\pi G\chi(\dot h - \Psi) = 0 \ ; \\
\dot\Delta + 9(\frac{\dot a}{a})^2(\frac{\chi}{\rho} - \chi')\Delta +
(1 + \alpha - 6\frac{\chi}{\rho}\frac{\dot a}{a})(\Psi - \dot h) = 0 \ ; \\
(1 + \alpha - 3\frac{\dot a}{a}\frac{\chi}{\rho})\dot\Psi +
\biggr[(1 + \alpha)(2 - 3\alpha)\frac{\dot a}{a} - 3\biggr(\frac{\ddot
a}{a} +(\frac{\dot a}{a})^2\biggl) + 9\alpha(\frac{\dot
a}{a})^2\frac{\chi}{\rho} +\nonumber \\
9\chi'(\frac{\dot a}{a})^2\biggr((1 + \alpha) -
3\frac{\dot a}{a}\frac{\chi}{\rho}
\biggl)\biggl]\Psi - (\frac{q}{a})^2\biggr[\alpha\Delta - 3\chi'\frac{\dot
a}{a}\Delta - \frac{\chi}{\rho}(\Psi - \dot h)\biggl] = 0 \ ,
\end{eqnarray}
where $\Delta = \delta\rho/\rho$, $\Psi ={\delta^i}_{,i}$
and $\chi' = d\chi/d\rho$.\\
We will now determine the perturbed solutions for the density contrast.
If $\chi = \chi_{_0}$, $\chi' = 0$, then using the background relation
\begin{equation}
1 + \alpha =3(\chi/\rho)(\dot a/a)\nonumber \ ,
\end{equation}
we obtain
\begin{equation}
\Delta = e^{-3Ht} \ .
\end{equation}
This solution cannot be eliminated by a residual coordinate transformation.
So, if the viscosity coefficient is constant, we have a de Sitter phase,
during which the density perturbations do not vanish; however they decrease
exponentially with time.\\
Let us now consider the case $\chi = \chi_{_0}\rho$. In the asymptotic limit
$t \rightarrow - \infty$, the density contrast decays exponentially as the
inverse of the cubic power of the scale factor at that stage. We find again
a de Sitter phase, with non-vanishing density perturbations, which however
decrease exponentially with time. On the other hand, in the limit $t
\rightarrow \infty$, the perturbed solutions approach the ones found in the
case of a perfect fluid, like the background solutions do.
\vspace{0.5cm}
\section{Residual gauge freedom}
\vspace{0.5cm}
We consider the infinitesimal coordinate transformation
$\tilde x^A \rightarrow x^A + \chi^A$, under which
the components of the metric tensor transform to
\begin{equation}
\tilde g_{AB} = g_{AB} + \chi_{(A;B)} \ .
\end{equation}
Imposing that the synchronous coordinate condition be preserved, we 
obtain the
following solutions for the time- and space-components  (in each of the two
subspaces) of $\chi^A$\ :
\begin{eqnarray}
\chi^0 &=& \Psi(x) \ , \\
\chi^i &=& \Psi(x)^{,i}\int{\frac{dt}{a^2}} + {\zeta(x)}^{,i} \ , \\
\chi^p &=& \Psi(x)^{,a}\int{\frac{dt}{b^2}} + {\Xi(x)}^{,a} \ ,
\end{eqnarray}
where $\Psi $, $\zeta $ and $\Xi $ are arbitrary functions. Since the traces
of the perturbed metric tensor, in each subspace, are
\begin{eqnarray}
\tilde h_k^{\ k} &=& h_k^{\ k} - 
a^2\biggr[2\Psi(x)_{,k,k}\int{\frac{dt}{a^2}}
+ 2\zeta_{,k,k} - 2d_1\frac{\dot a}{a}\Psi\biggl] \ , \\
\tilde h_a^{\ a} &=& h_a^{\ a}-b^2\biggr[2\Psi(x)_{,a,a} \int{\frac{dt}{b^2}}
+ 2\Xi_{,a,a} - 2d_2\frac{\dot b}{b}\Psi\biggl] \ ,
\end{eqnarray}
one can show, using the definitions $h = h_k^{\ k}/a^2$, $H = h_a^{\ a}/b^2$
and the equations for the perturbations, that the density contrast transforms
under this residual coordinate transformation as
\begin{equation}
\tilde\Delta = \Delta - \left [(1 + \alpha_1)d_1\frac{\dot a}{a} +
 (1 + \alpha_2)d_2\frac{\dot b}{b}\right ]\Psi(x) \ .
\end{equation}
Thus, any perturbed solution with a time behaviour
$(1+\alpha_1)d_1(\dot a/a) + (1+\alpha_2)d_2(\dot b/b)$
can in principle be eliminated by a coordinate transformation and has no
physical meaning.
\vspace{1cm}\\
\end{appendix}
{\bf Acknowledgement}
\vspace{0.5cm}\\
It is a pleasure to thank J\'er\^ome Martin, for his participation during the
beginning of this project and for a number of helpful discussions.  We want
also to thank the Laboratoire de Gravitation et Cosmologie Relativistes,
Universit\'e Pierre et Marie Curie, where a part of this work was done.
One of us (M.S.) wants to thank the Departamento de F\'{\i}sica, Universidade
Federal do Esp\'{\i}rito Santo for warm hospitality during the last 
stages of
this work.

\end{document}